# Evolution of graphene growth on Cu and Ni studied by carbon isotope labeling

Xuesong Li[a], Weiwei Cai[a], Luigi Colombo[b*], and Rodney S. Ruoff[a*]

**Large-area graphene is a new material with properties that make it desirable for advanced scaled electronic devices[1]. Recently, chemical vapor deposition (CVD) of graphene and few-layer graphene using hydrocarbons on metal substrates such as Ni and Cu has shown to be a promising technique[2-5]. It has been proposed in recent publications that graphene growth on Ni occurs by C segregation[2] or precipitation[3], while that on Cu is by surface adsorption[5]. In this letter, we used a carbon isotope labeling technique to elucidate the growth kinetics and unambiguously demonstrate that graphene growth on Cu is by surface adsorption whereas on Ni is by segregation-precipitation. An understanding of the evolution of graphene growth and thus growth mechanism(s) is desired to obtain uniform graphene films. The results presented in this letter clearly demonstrate that surface adsorption is preferred over precipitation to grow graphene because it is a self-limiting process and thus manufacturable.**

Graphene, a monolayer of $sp^2$-bonded carbon atoms or one monolayer of graphite, has attracted interest in part because of its unique transport properties[1]. The surface science community has an extensive literature on what was referred to as "monolayer graphite", i.e., graphene, as grown on various metal films that are epitaxially well matched to graphene[6]. Recent attempts to obtain graphene, including on insulating substrates for device measurements, have been by chemical reduction of exfoliated graphite oxide layers and subsequent deposition from colloidal suspensions[7-10], by ultrahigh vacuum (UHV) annealing of single crystal SiC[11,12], and by growth on metal substrates[2-5,13-15]. To date it has not been possible to fully recover the electronic properties of graphene in reduced exfoliated graphite oxide layers. Graphene obtained on SiC single crystals has shown good transport properties, but this may be limited to devices on SiC since transfer to other substrates such as $SiO_2$/Si has not been demonstrated yet and might be difficult. There have been a number of reports on the growth of graphene on metal substrates such as Ni, Co, Ru, Ir, Cu, etc. with the use of UHV[13-15] or 'normal' CVD[2-5] systems. Growth of graphene and FLG on polycrystalline Ni and Cu substrates by CVD has the advantage of producing continuous large-area films. It has been proposed that CVD growth of graphene on Ni is due to a C segregation[2] or precipitation[3] process and a fast cooling rate is suggested as critical for suppressing formation of multiple layers and thus obtaining graphene or FLG[4]. The graphene films grown on Ni foil/film so far however are still not uniformly monolayer, i.e., they have a wide variation in thickness over the film area. Recently, we have shown that Cu is an excellent candidate for making large-area graphene films with uniform thickness due to the low solubility of C in Cu[5]. We also proposed a surface adsorption growth mechanism for this case, namely a self-limiting process that 'automatically' yields graphene instead of multilayer material with poor control of thickness.

---

[a] Department of Mechanical Engineering and the Texas Materials Institute, 1 University Station C2200, The University of Texas at Austin, Austin, TX 78712-0292
[b] Texas Instruments Incorporated, Dallas, TX 75243
*Corresponding authors: r.ruoff@mail.utexas.edu, colombo@ti.com

We used isotopic labeling of the carbon precursor to study the mechanism and kinetics of CVD growth and of graphene and graphite on Cu and Ni substrates. Cu foils (25-μm thick) and 700-nm thick Ni films deposited on $SiO_2$/Si substrates by sputtering were used as the metal substrates. The experimental procedure for graphene growth was similar to that reported previously[5] with a deposition temperature ranging from 900-1000 ºC. However, in this work both normal methane and $^{13}CH_4$ (99.95% pure) were introduced to the growth chamber in particular sequences. The duration of exposure of methane is defined as $^jt_i$ as, where j = 12 or 13 denotes $^{12}CH_4$ or $^{13}CH_4$, and i denotes the step in the sequence (e.g., $^{13}t_1$ means the first gas introduced was $^{13}CH_4$ with the duration of exposure being $t_1$).

We took advantage of the separation of the $^{12}C$ and $^{13}C$ Raman modes to observe the spatial distribution of graphene domains. The frequencies of Raman modes are given by Eq (1) with the assumption that the $^{12}C$ or $^{13}C$ atoms are randomly mixed and the bond force constants are equal[16]:

$$\omega = \omega_{12}\sqrt{\frac{m_{12}}{n_{12}m_{12} + n_{13}m_{13}}} \qquad (1)$$

The Raman mode frequency of $^{12}C$ graphene/graphite is $\omega_{12}$, $n_{12}$ and $n_{13}$ are the atomic fractions and $m_{12}$ and $m_{13}$ are the atomic masses of $^{12}C$ and $^{13}C$, respectively.

Fig. 1 shows the possible distribution of $^{12}C$ and $^{13}C$ in the graphene films based on different assumed growth mechanisms when $^{12}CH_4$ or $^{13}CH_4$ are introduced sequentially. Fig. 1a shows the case of (segregation)-precipitation growth. *Segregation* and *precipitation* are different concepts, though both show compositional heterogeneity at lattice discontinuities[17]. Some metals such as Ni show segregation of C[17,18] while others such as Pt (110) do not[19]. For the case of sequential dosing of $^{12}CH_4$ and $^{13}CH_4$ yielding formation of a C-metal solution, the segregated and/or precipitated graphene will consist of randomly mixed isotopes. In contrast, if graphene with the sequential dosing of $^{12}CH_4$ and $^{13}CH_4$ grows by surface adsorption, the isotope distribution in the local graphene regions will reflect the dosing sequence employed (Fig. 1b). The *equilibrium adsorption* and *equilibrium segregation* at a solid vapor interface in a binary system are thermodynamically identical and only the source of the adsorbate is different, the source being from either the gas phase or from the bulk, respectively[17]. It is also possible to have a combination of *surface adsorption* and *precipitation*. In this case, the top layer (which grows first at the deposition temperature and is obtained by surface adsorption) will consist of regions that are $^{12}C$-pure and regions that are $^{13}C$-pure (again, reflecting the particular dosing protocol that was used).

Layer(s) below the top layer then result from precipitation of C from the bulk solution and would show as a uniform distribution of $^{12}$C and $^{13}$C (Fig. 1c).

Fig. 2 shows the results of graphene grown on Cu with a deposition temperature of 1000 °C where the carbon precursor was introduced according to the following sequence: $^{13}t_{1,3,5,7} = {}^{12}t_{2,4,6,8} = 1$ min. Fig. 2a shows an optical micrograph of the resulting graphene film transferred onto a SiO$_2$/Si substrate using polymethyl-methacrylate (PMMA) as the carrier material for transfer as previously reported[5,20]. The surface is relatively uniform with the exception of wrinkles formed during cool-down that occur due to the different coefficients of thermal expansion between graphene/graphite and the underlying metal substrate[21,22]. The Raman spectra, Fig. 2b, show a graphene film with regions having close to pure $^{12}$C from natural methane (~99% $^{12}$CH$_4$), regions of isotopically pure $^{13}$C, and regions where both $^{12}$C and $^{13}$C are present. An analysis of both the color contrast of the optical micrograph[23] and the Raman spectra[3,24-26] show that the carbon layer is one monolayer graphite or graphene. When the Raman laser beam was focused at the junction of $^{12}$C and $^{13}$C graphene regions the characteristic bands for both $^{12}$C and $^{13}$C graphene appeared in the spectrum. The intensity of each band depends on the area occupied by each isotopically-labeled region under the laser spot and the sum of the intensity of two bands (e.g., G$^{13}$+G$^{12}$) was found to be essentially equal to that of the intensity from either the pure $^{13}$C or $^{12}$C regions for a graphene film.

The films were also analyzed using a Raman spectroscopy mapping technique to identify the spatial distribution of $^{12}$C and $^{13}$C. Figs. 2d-i display Raman G and D band maps of the film shown in Fig. 2a for both $^{12}$C and $^{13}$C. Fig. 2d is a map of the overall G band intensity (G$^{13}$ + G$^{12}$) of the area shown in Fig. 2a. The uniform intensity distribution proves good thickness uniformity except for the wrinkles (bright lines). Figs. 2e & f are the maps of the G-band of $^{13}$C and of $^{12}$C, respectively, which show the time evolution of graphene growth. The bright solid centers in the G$^{13}$ map in Fig. 2e correspond to $^{13}$C-graphene grown during $^{13}t_1$; the low intensity, dark rings correspond to $^{12}C$-graphene grown during $^{12}t_2$, which are seen as bright rings in the $G^{12}$ map in Fig. 2f; the bright area between the dark rings in Fig. 2e corresponds to $^{13}C$-graphene grown during $^{13}t_3$. Fig. 2c shows a line scan (marked with dashed lines across Figs. 2d-f) where the $^{12}C$-graphene and $^{13}C$-graphene domains are clearly seen with the blue line representing the $G^{13}$ (i.e., $^{13}C$-graphene) domains and the pink line representing the $G^{12}$ (i.e., $^{12}C$-graphene) domains. The green line, which is the most uniform across the film, is the overall G band intensity ($G^{13} + G^{12}$) with the peak corresponding to the wrinkle in the film. It is interesting to note that we flowed 8 cycles of alternating $^{12}CH_4$ and $^{13}CH_4$ but the resulting graphene grew only during the first three dosings. The 4$^{th}$ and subsequent doses played no role because the surface was already saturated with graphene. This data shows that single layer graphene on Cu grows in less than 3 minutes under the conditions we used, growth occurs in 2-dimensions (thus, is a consequence of a surface-adsorption process), and that growth is self-limited since there is no catalyst to promote decomposition and growth after the first layer of carbon (graphene) is deposited.

Further, the films grown on Cu show little to no detectable Raman D-band indicating that the graphene has few defects. In areas where a D band is observed, it may be attributed to the presence of edges from small bi-layers regions, domain boundaries, or defective centers at nucleation sites. The D maps in Figs. 2g-i provide additional information. Other than the high intensity of the D band from the wrinkles, some bright spots and lines are also shown corresponding to defective centers and boundaries between graphene domain, respectively. These inter-domain defects occur where the graphene domains join, for example in the current case after $^{13}t_3$ as indicated by the blue arrows in Fig. 2g. A possible reason for these defects could be the formation of pentagonal and/or heptagonal arrangements of carbon atoms that form mis-oriented graphene domains resulting from the Cu surface roughness, as was found for Ir[27]. In contrast, the low defect boundaries (indicated by the white arrows in Fig. 2g) may indicate a "good" connection between two domains. A detailed understanding of such defects is suggested for future work. Careful observation of the optical micrograph (Fig. 2a) and the Raman maps (Figs. 2d-i) shows that there is no overlap of the graphene layers where the domains join, suggesting that there is crystallographic registration to the Cu substrate. If there were overlap, a high contrast or bright line would be present in the micrograph and the G-band Raman maps. This sequential distribution of $^{13}$C and $^{12}$C clearly shows that graphene growth on Cu is based on the surface adsorption mechanism. We have not observed randomly distributed $^{12}$C and $^{13}$C on the many graphene films that we have grown on Cu indicating that graphene does not grow by a precipitation process from the bulk. This may be attributed to the extremely low solubility of C in Cu.

The graphene growth rate can be derived from the timing sequence of the $^{12}$CH$_4$ and $^{13}$CH$_4$ flow and Raman imaging data. The *area growth rate* (the increase in area of individual grains per unit time) is 36 μm$^2$/min in an average and the *edge growth rate* (the advance of the grain edge per unit time) is 1~6 μm/min, which may be graphene/Cu orientation dependant.

Fig. 3 shows the results of graphene growth on 700-nm thick Ni film with a deposition temperature of 900 °C. We evaluated several sets of pre-designed feeding time/sequences of isotopes, but no distinguishable separation of isotopes was found. The fraction of isotopes in the film only depended on the feeding time/sequences in the first several minutes of the deposition time. Fig. 3a shows the fraction of $^{13}$C as a function of $^{13}$t$_1$ while $^{12}$t$_2$ = 10 min was kept constant. It can be seen that between 3 and 4 min there is a transition which can be attributed to the formation of a graphene layer on the Ni film due to surface segregation[17,18]. When the feeding time, $^{13}$t$_1$ is less than 3 min, the Ni film is not saturated with carbon. After the exposure to $^{13}$CH$_4$ is stopped, the Ni surface is still "open" and as a result the pre-dissolved $^{13}$C atoms are displaced by the $^{12}$C atoms supplied afterwards, and the final film consists mainly of $^{12}$C. As the exposure time is increased, the C concentration in the Ni film increases and formation of graphene begins as a result of surface segregation. Switching to $^{12}CH_4$ at this transition period, e.g., $^{13}t_1$ = 3.5 min, produced graphene with mixed $^{12}C$ and $^{13}C$. When the feeding time, $^{13}t_1$, is long enough, e.g., greater than 4 min, the Ni surface shows full coverage by the segregated graphene. This graphene layer is stable and stops the gas ($CH_4$, $H_2$) from reacting with the solid Ni and exposure to more $^{12}CH_4$ has no effect on the final product. The large error bars near the transition region suggest that the distribution of isotopes is not laterally uniform because the Ni film is not saturated and the microstructure may affect the segregation process.

Fig. 3c shows an optical micrograph of a graphene film grown on a Ni film ($^{13}t_{1,3,5,7}$ = $^{12}t_{2,4,6,8}$ = 1 min) and transferred onto a $SiO_2$/Si wafer by PMMA similar to the reported method[5,20]. Compared to graphene grown on Cu foil, the film on Ni is not uniform in thickness and it consists of 1 to tens of graphene layers across the Ni surface. The growth of extra graphene layers can be attributed to C precipitation during the cooling process as verified by the random distribution of the two C isotopes as measured by Raman imaging and shown in Fig. 3d. The D map in Fig. 3e shows that the graphene film has a low defect density.

The two mechanisms of graphene growth on Cu and Ni can be understood from the C-metal binary phase diagram, which is schematically shown in Fig. 4. The binary phase diagrams of C-Cu and C-Ni are similar in that C has a limited solubility in the metal without the presence of a metal-carbide line compound. The only significant difference is that the solubility of C in Cu is *much* lower than that in Ni. In Fig.4, $[C]_p$ corresponds to the solubility of C in metal and $[C]_s$ corresponds to the concentration for surface segregation or adsorption, which is less than $[C]_p$ for the case of Ni. Although we don't know if there is C surface segregation for the C-Cu system, we can set $[C]_s = [C]_p$ if there is not, since it will not affect our discussion. Under isothermal and isobaric process conditions, the concentration of C in the metal, $[C]$, increases with deposition time and saturates finally when $[C] = [C]_p$. However, surface segregation or adsorption may start once $[C] > [C]_s$, and by controlling the deposition time, we can control $[C]$. When $[C] < [C]_s$ at $T_d$, e.g., $[C] = [C]_1$, there is no graphene growth at $T_d$. When T decreases to $T_{s1}$, where $[C]_s = [C]_1$, graphene will grow due to surface segregation. When T decreases to $T_{p1}$, more C will precipitate out due to the decrease of solubility, as shown by the red curve in Fig. 4b. For example, graphene grew on 700-nm Ni film with $T_d = 900\ ^oC$ and a deposition time less than 3 min. If the deposition time is longer than 4 min so that $[C] > [C]_s$, e.g., $[C] = [C]_2$, graphene will grow at $T_d$. The C atoms in the graphene layer are mainly from the bulk of the Ni film since the solubility of C in Ni is high. In contrast, the solubility of C in Cu is very small so that the region near the surface can be saturated quickly and the C for graphene formation is from the gas phase (adsorption). When T decreases to $T_{p2}$, more C will precipitate out (blue curve in Fig. 4b) from the Ni film but there is no evidence of this from the Cu foil.

The surface adsorption process at $T_d$ is an equilibrium process and only one graphene layer grows on the metal as a result. In this process, carbon adsorption is surface mediated and growth ends once the surface is fully covered with graphene. Since the carbon source is from the vapor phase and not from the bulk of the metal and the graphene surface is chemically inert a monolayer graphite is ensured. In contrast, the C precipitation process is a non-equilibrium process, which should be suppressed if one aims to achieve graphene growth, as is the case for Ni-C[2-4]. However, because of microstructural defects it is very difficult to fully eliminate the effect of precipitation for metals with high carbon solubility, especially when the metal grain size is very small since more carbon tends to precipitate out at the grain boundaries, which will result in a non-uniform growth. Hence, metals with low C solubility such as Cu offer a better path to large-area growth of graphene. One of the principal issues in the manufacturing of thin films of any material is thickness and composition control. In the case of graphene this would be an even more difficult problem since a single layer of graphite would have to be grown. The data presented in this paper on graphene growth on Cu substrates clearly marks a path to monolayer graphite growth on Cu and perhaps also on other substrate materials where carbon solubility is low and precipitation is not observed. However, if one wants to grow controlled bi-layer graphene which has been found to show some desirable electronic properties, then a precipitation process will have to be taken advantage of in conjunction with solubility[28]. Since Cu and Ni are miscible, one can envision a Cu-Ni alloy with appropriate composition to tune the solubility of C in the alloy[29] and hence enable bilayer graphene growth (or trilayer, etc.).

**Acknowledgement**. We would like to thank the Nanoelectronic Research Initiative (NRI-SWAN; #2006-NE-1464), DARPA under contract FA8650-08-C-7838 through the CERA program, and The University of Texas at Austin for support.

**Figure captions.**

**Figure 1. Schematics of the possible distribution of C isotopes in the graphene films based on different growth mechanisms for sequential input of C isotopes.** (a) Graphene with randomly mixed isotopes such as might occur from surface segregation and/or precipitation. (b) Graphene with separated isotopes such as might occur by surface adsorption. (c) Combined growth from surface adsorption and precipitation.

**Figure 2. Micro-Raman characterization of the isotope-labeled graphene grown on Cu foil and transferred onto a SiO$_2$/Si wafer.** (a) An optical micrograph of the identical region analyzed with micro-Raman spectroscopy. (b) Raman spectra from $^{12}$C-graphene (green), $^{13}$C-graphene (blue), and the junction of $^{12}$C- and $^{13}$C-graphene (red), respectively, marked with the corresponded colored circles in (a) and (e). (c) Line scan of the dashed lines in (d-f). Raman maps of (d) G$^{13+12}$(1500-1620 cm$^{-1}$), (e) G$^{13}$(1500-1560 cm$^{-1}$), (f) G$^{12}$(1560-1620 cm$^{-1}$), (g) D$^{13+12}$(1275-1375 cm$^{-1}$), (h) D$^{13}$(1275-1325 cm$^{-1}$), and (i) D$^{12}$(1325-1375 cm$^{-1}$), of the area shown in (a). Scale bars are 5 μm.

**Figure 3. Distribution of C isotopes in a FLG film grown on Ni.** (a) Fraction of $^{13}$C as a function of $^{13}$t$_1$ with $^{12}$t$_2$ fixed at 10 min. The inset shows three Raman spectra of FLG films on Ni each having a different isotope composition. (b) An optical micrograph of a FLG film transferred onto a SiO$_2$/Si wafer and the corresponding Raman maps of (c) G (1526-1586 cm$^{-1}$) and (d) D (1305-1355 cm$^{-1}$), of the same region, showing the film consists of randomly mixed isotopes (with an overall composition of ~40% $^{13}$C and ~60% $^{12}$C) with low defect density. Scale bars are 5 μm.

**Figure 4. Interpretation of different growth mechanisms associated with the C/metal binary phase diagram.** (a) Schematic C/metal binary phase diagram. (b) The number of graphene layers as a function of temperature.

**Figures**

Figure 1

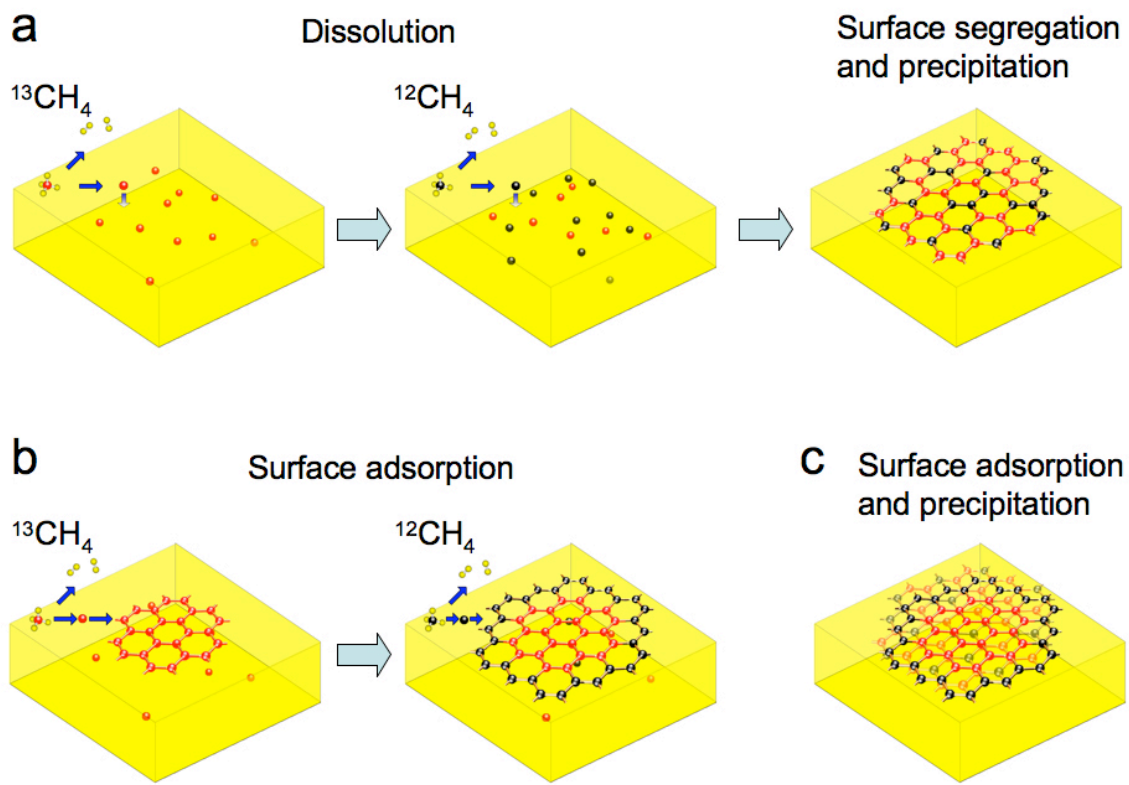

Figure 2.

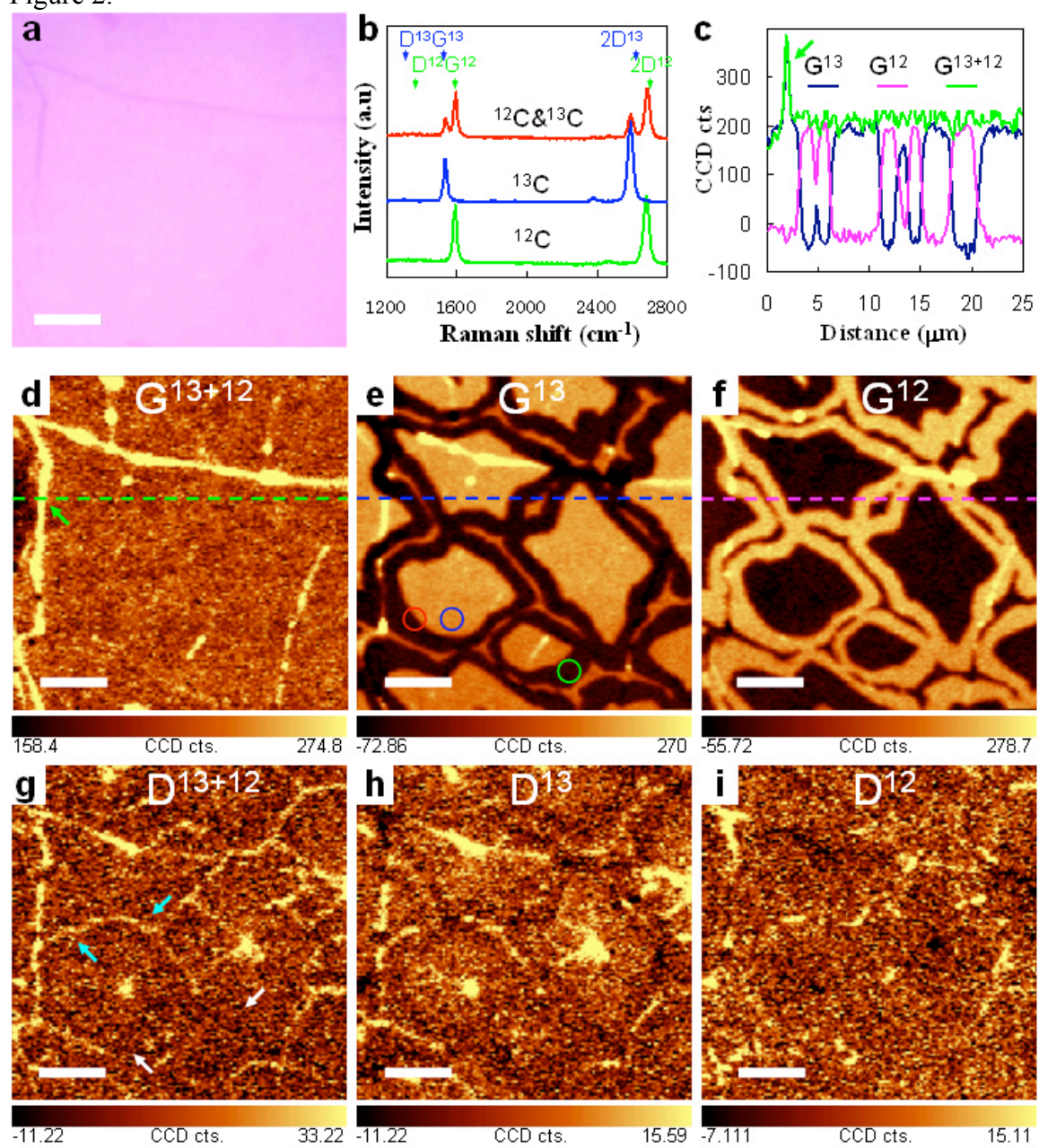

Figure 3

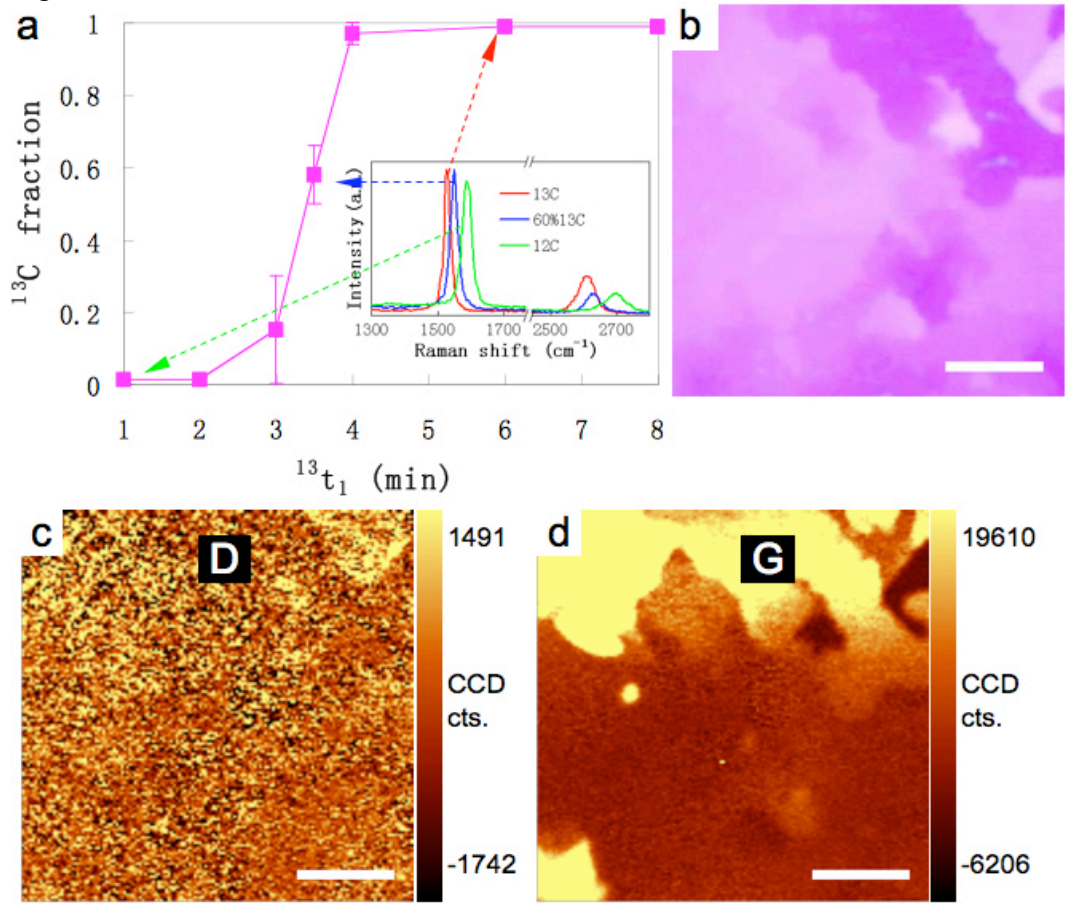

Figure 4

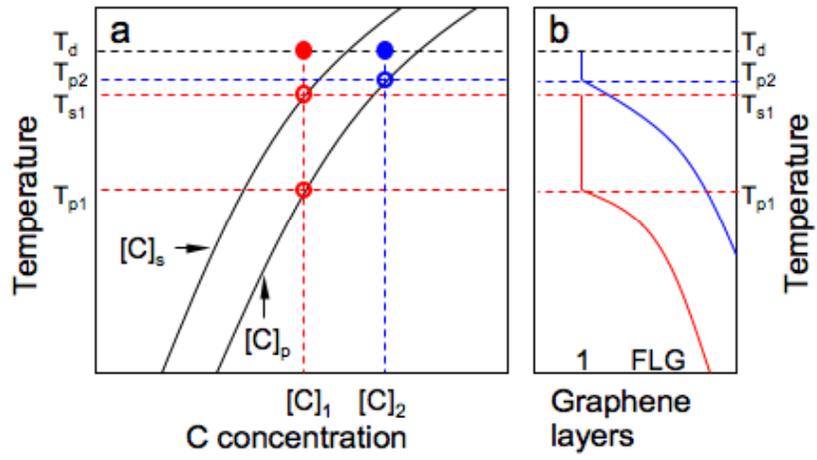